
\newif\ifPubSub \PubSubfalse
\newcount\PubSubMag \PubSubMag=1200
\def\PubSub{\PubSubtrue
            \magnification=\PubSubMag \hoffset=0pt \voffset=0pt
            \pretolerance=600 \tolerance=1200 \vbadness=1000 \hfuzz=2 true pt
            \baselineskip=1.75\baselineskip plus.2pt minus.1pt
            \parskip=2pt plus 6pt
            \setbox\strutbox=\hbox{\vrule height .75\baselineskip
                                               depth  .25\baselineskip
                                               width 0pt}%
            \Page{5.75 true in}{8.9 true in}}

\newcount\FigNo \FigNo=0
\newbox\CapBox
\newbox\FigBox
\newtoks\0
\def\Fig {fig.\thinspace\the\FigNo}
\def\NFig {{\count255=\FigNo \advance\count255 by 1 fig.\thinspace\the\count255}}
\def\StartFigure #1#2#3{\global\advance\FigNo by 1
                  \ifPubSub \global\setbox\FigBox=\vbox\bgroup
                            \unvbox\FigBox
                            \parindent=0pt \parskip0pt
                            \eject \line{\hfil}\bigskip\bigskip
                  \else \midinsert \removelastskip \bigskip\bigskip
                  \fi
                  \begingroup \hfuzz1in
                  \dimen0=\hsize \advance\dimen0 by-2\parindent \indent
                  \vbox\bgroup\hsize=\dimen0 \parindent=0pt \parskip=0pt
                       \vrule height1pt depth0pt width0pt
                       \ifdim\dimen0<#1bp \dimen1=#1bp
                                          \advance\dimen1 by -\dimen0
                                          \divide\dimen1 by 2 \hskip-\dimen1
                       \fi
                       \hfil
                       \vbox to #2 bp\bgroup\hsize #1 bp
                            \vss\noindent\strut\special{"#3}%
                            \skip0=\parskip \advance\skip0 by\dp\strutbox
                            \vskip-\skip0 }%
\def\caption #1{\strut\egroup
                \ifPubSub \global\setbox\CapBox=\vbox{\unvbox\CapBox
                                        \parindent0pt \medskip
                                         {\bf Figure \the\FigNo:}
                                          {\tenrm\strut #1\strut}}%
                \else \bigskip {\FootCapFace \hfuzz=1pt
                                \noindent{\bf Figure~\the\FigNo:} #1\par}%
                \fi}%
\def\label (#1,#2)#3{{\offinterlineskip \parindent=0pt \parskip=0pt
                     \vskip-\parskip 
                     \vbox to 0pt{\vss
                          \moveright #1bp\hbox to 0pt{\raise #2bp
                                         \hbox{#3}\hss}\hskip-#1bp\relax}}}%
\def\EndFigure {\egroup \endgroup 
                \ifPubSub \vfil
                          \centerline{\bf Figure \the\FigNo}
                          \egroup 
                \else \bigskip \smallskip \endinsert 
                \fi }
\def\ListCaptions {\vfil\eject \message{!Figure captions:!}%
                   \Sectionvar{Figure Captions}\par
                   \unvbox\CapBox}
\def\ShowFigures {\ifPubSub\ifnum\FigNo>0 \ListCaptions \vfil\eject
                                 \message{!Figures:!}%
                                 \nopagenumbers \unvbox\FigBox \eject
                           \fi
                  \fi}
\def\FootCapFace{} 


\parskip=0pt plus 3pt
\def\Page #1#2{{\dimen0=\hsize \advance\dimen0 by-#1 \divide\dimen0 by 2
               \global\advance\hoffset by \dimen0
               \dimen0=\vsize \advance\dimen0 by-#2 \divide\dimen0 by 4
               \ifdim\dimen0<0pt \multiply\dimen0 by 3 \fi
               \global\advance\voffset by \dimen0
               \global\hsize=#1 \global\vsize=#2\relax}
               \ifdim\hsize<5.5in \tolerance=300 \pretolerance=300 \fi}
\Page{5in}{8in} 
\def\EndPaper{\par\dosupereject \ifnum\RefNo>1 \ShowReferences \fi
              \ShowFigures
              \par\vfill\supereject
              \message{!!That's it.}\end}
\headline{\ifnum\pageno=-1 \hfil \Smallrm \Time, \Date \else \hfil \fi}
\def\VersionInfo #1{\headline{\ifnum\pageno=-1 \hfil \Smallrm #1
                              \else \hfil
                              \fi}}


\newcount\RefNo \RefNo=1
\newbox\RefBox
\def\Jou #1{{\it #1}}
\def\Vol #1{{\bf #1}}
\def\AddRef #1{\setbox\RefBox=\vbox{\unvbox\RefBox
                       \parindent1.75em
                       \pretolerance=1000 \hbadness=1000
                       \vskip0pt plus1pt
                       \item{\the\RefNo.}
                       \sfcode`\.=1000 \strut#1\strut}%
               \global\advance\RefNo by1 }
\def\Ref  #1{\Hunskip~[{\the\RefNo}]\AddRef{#1}}
\def\Refc #1{\Hunskip~[{\the\RefNo,$\,$}\AddRef{#1}}
\def\Refm #1{\Hunskip{\the\RefNo,$\,$}\AddRef{#1}}
\def\Refe #1{\Hunskip{\the\RefNo}]\AddRef{#1}}
\def\Refl #1{\Hunskip~[{\the\RefNo}--\AddRef{#1}}
\def\Refn #1{\Hunskip\AddRef{#1}}
\def\ShowReferences {\message{!References:!}%
                     \Sectionvar{References}\par
                     \unvbox\RefBox}
\def\StoreRef #1{\Hunskip\edef#1{\the\RefNo}}


\newif\ifRomanNum
\newcount \Sno \Sno=0
\def\Interskip #1#2#3{{\removelastskip \dimen0=#1
                       \advance\dimen0 by2\baselineskip
                       \vskip0pt plus\dimen0 \penalty-300
                       \vskip0pt plus-\dimen0
                       \advance\dimen0 by -2\baselineskip
                       \vskip\dimen0 plus#2 minus#3}}
\def\Section #1\par{\Interskip{24pt}{6pt}{2pt}%
                    \global\advance\Sno by1
                    \setbox0=\hbox{\STitlefont
                                    \ifRomanNum
                                       \global\SSno=64 \uppercase
                                       \expandafter{\romannumeral
                                                    \the\Sno.\ \ }%
                                     \else
                                        \global\SSno=96 \the\Sno.\ \
                                     \fi}
                    \leftline{\vtop{\copy0 }%
                              \vtop{\advance\hsize by -\wd0
                                    \raggedright
                                    \pretolerance10000 \hbadness10000
                                    \noindent \STitlefont
                                    \GetParenDim{\dimen0}{\dimen1}%
                                    \advance\dimen0 by\dimen1
                                    \baselineskip=\dimen0 #1}}
                    \hrule height0pt depth0pt
                    \dimen0=\baselineskip \advance\dimen0 by -\parskip
                    \nobreak\vskip\dimen0 plus 3pt minus3pt \noindent}%
\def\Sectionvar #1\par{\Interskip{24pt}{6pt}{2pt}%
                    \leftline{\vbox{\noindent
                                    \STitlefont #1}}
                    \hrule height0pt depth0pt
                    \dimen0=\baselineskip \advance\dimen0 by -\parskip
                    \nobreak\vskip\dimen0 plus 3pt minus3pt \noindent}%
\newcount\SSno \SSno=0
\def\SubSection #1\par{\Interskip{15pt}{3pt}{1pt}%
                    \global\advance\SSno by1
                    \setbox0=\hbox{\SSTitlefont \char\SSno$\,$]$\,$\ }
                    \leftline{\vtop{\copy0 }%
                              \vtop{\advance\hsize by -\wd0
                                    \raggedright
                                    \pretolerance10000 \hbadness10000
                                    \noindent \SSTitlefont
                                    \GetParenDim{\dimen0}{\dimen1}%
                                    \advance\dimen0 by\dimen1
                                    \baselineskip=\dimen0 #1}}
                    \hrule height0pt depth0pt
                    \dimen0=.8\baselineskip \advance\dimen0 by -2\parskip
                    \nobreak\vskip\dimen0 plus1pt minus1pt \noindent}%
\def\(#1){~\ifRomanNum {\Medrm\uppercase\expandafter{\romannumeral #1}}%
           \else {#1}%
           \fi}


\newcount\EqNo \EqNo=0
\def\NumbEq {\global\advance\EqNo by 1
             \eqno(\the\EqNo)}
\def\PrevEq {(\the\EqNo)}
\def\PrevEqs #1{{\count255=\EqNo \advance\count255 by-#1\relax
                 (\the\count255)}}
\def\NameEq #1{\xdef#1{(\the\EqNo)}}

\def\AppndEq #1{\EqNo=0
  \def\NumbEq {\global\advance\EqNo by 1
             \eqno(#1\the\EqNo)}
  \def\numbeq {\global\advance\EqNo by 1
             (#1\the\EqNo)}
  \def\PrevEq {(#1\the\EqNo)}
  \def\PrevEqs ##1{{\count255=\EqNo \advance\count255 by-##1\relax
                 (#1\the\count255)}}
  \def\NameEq ##1{\xdef##1{(#1\the\EqNo)}}}


\newcount\ThmNo \ThmNo=0

\def\Theorem #1\par{\removelastskip\bigbreak
    \advance\ThmNo by 1
    \noindent{\bf Theorem \the\ThmNo:} {\sl #1\bigskip}}
\def\TheoremN #1\par{\removelastskip\bigbreak
    \noindent{\bf Theorem:} {\sl #1\bigskip}}
\def\Lemma #1\par{\removelastskip\bigbreak
    \advance\ThmNo by 1
    \noindent{\bf Lemma \the\ThmNo:} {\sl #1\bigskip}}

\def\PrevThm {Theorem~\the\ThmNo}
\def\PrevLm {lemma~\the\ThmNo}
\def\Declare #1#2\par{\removelastskip\bigbreak
    \noindent{\bf #1:} {\sl #2\bigskip}}
\def\Proof {\removelastskip\bigskip
    \noindent{\bf Proof:\ \thinspace}}
\def\EndProof{{~\nobreak\hfil \copy\Tombstone
               \parfillskip=0pt \bigskip}}


\newlinechar=`\!
\def\\{\ifhmode\hfil\break\fi}
\let\thsp=\,
\def\,{\ifmmode\thsp\else,\thinspace\fi}
\def\Hunskip {\ifhmode\unskip\fi}

\def\Date {\ifcase\month\or January\or February\or March\or April\or
 May\or June\or July\or August\or September\or October\or November\or
 December\fi\ \number\day, \number\year}

\newcount\mins  \newcount\hours
\def\Time{\hours=\time \mins=\time
     \divide\hours by60 \multiply\hours by60 \advance\mins by-\hours
     \divide\hours by60         
     \ifnum\hours=12 12:\ifnum\mins<10 0\fi\number\mins~P.M.\else
       \ifnum\hours>12 \advance\hours by-12 
         \number\hours:\ifnum\mins<10 0\fi\number\mins~P.M.\else
            \ifnum\hours=0 \hours=12 \fi
         \number\hours:\ifnum\mins<10 0\fi\number\mins~A.M.\fi
     \fi }

\def\HollowBox #1#2#3{{\dimen0=#1
       \advance\dimen0 by -#3 \dimen1=\dimen0 \advance\dimen1 by -#3
        \vrule height #1 depth #3 width #3
        \hskip -#3
        \vrule height 0pt depth #3 width #2
        \llap{\vrule height #1 depth -\dimen0 width #2}%
       \hskip -#3
       \vrule height #1 depth #3 width #3}}
\newbox\Tombstone
\setbox\Tombstone=\vbox{\hbox{\HollowBox{8pt}{5pt}{.8pt}}}

\def\GetParenDim #1#2{\setbox1=\hbox{(}%
                      #1=\ht1 \advance#1 by 1pt
                      #2=\dp1 \advance#2 by 1pt}

\def\Bull #1#2\par{\removelastskip\smallbreak\textindent{$\bullet$}
                    {\it #1}#2\smallskip}
\def\Heading #1#2\par{\removelastskip\smallbreak\noindent {\it #1.}%
                      \nobreak\par #2\smallskip}


\def\Title #1\par{\message{!!#1}%
                  \nopagenumbers \pageno=-1
                  \null \bigskip
                  {\leftskip=0pt plus 1fil \rightskip=\leftskip
                   \parfillskip=0pt\relax \Titlefont
                   \ifdim\baselineskip<2.5ex \baselineskip=2.5ex \fi
                   \noindent#1\par}\bigskip}
\def\Author #1\par{{\bigskip
                     \count1=0   
                     \count2=0   
                     \dimen0=0pt 
                    \Position{#1\hskip-\dimen0 }\bigskip}}
\def\Address #1\par{{\leftskip=\parindent plus 1fil \rightskip=\leftskip
                     \parfillskip=0pt\relax
                     \ifdim\baselineskip>3ex \baselineskip=3ex \fi
                     \def\\{\break}
                     \noindent#1\par}\bigskip}
\def\modfnote #1#2{{\parindent=1.1em \leftskip=0pt \rightskip=0pt
                    \GetParenDim{\dimen1}{\dimen2}%
                    \setbox0=\hbox{\vrule height\dimen1 depth\dimen2
                                          width 0pt}%
                    \advance\dimen1 by \dimen2 \baselineskip=\dimen1
                    \vfootnote{#1}{\hangindent=\parindent \hangafter=1
                                 \unhcopy0 #2\unhbox0
                                 \vskip-\baselineskip \vskip1pt}}}%
\def\PAddress #1{\ifcase\count1 \let\symbol=\dag
                 \or \let\symbol=\ddag
                 \or \let\symbol=\P
                 \or \let\symbol=\S
                 \else \advance\count1 by -3
                       \def\symbol{\dag_{\the\count1 }}%
                 \fi
                 \advance\count1 by 1
                 \setbox0=\hbox{$^{\symbol}$}\advance\dimen0 by \wd0
                 \Hunskip \box0
                 \modfnote{$\symbol$}{{\sl Permanent address\/}: #1}}%
\def\Email #1{\ifcase\count2 \let\symbol=\ast
                 \or \let\symbol=\star
                 \or \let\symbol=\bullet
                 \or \let\symbol=\diamond
                 \or \let\symbol=\circ
                 \else \advance\count2 by -4
                       \def\symbol{\ast_{\the\count2 }}%
                 \fi
                 \advance\count2 by 1
                 \setbox0=\hbox{$^{\textstyle\symbol}$}\advance\dimen0 by \wd0
                 \Hunskip \box0
                 \modfnote{$\symbol$}{{\sl Electronic mail\/}: #1}}%
\def\Abstract{\vfil \message{!Abstract:!}%
              \Position{{\STitlefont Abstract}}
              \medskip \bgroup
              \ifdim\baselineskip>3.5ex \baselineskip=3.5ex \fi
              \narrower\noindent\ignorespaces}%
\def\EndAbstract{\par\egroup}
\def\pacs #1{\vfil\leftline{PACS numbers: #1}\eject}
\def\StartPaper{\message{!Body:!}%
                \pageno=1 \ifPubSub \footline{\tenrm\hss
                                              --\ \folio\ -- \hss}%
                          \else \footline{\hss\vtop to 0pt{\hsize=.15\hsize
                                     \vglue12pt \hrule \medskip
                                     \centerline{\tenrm\folio}\vss}\hss}%
                          \fi}


\RomanNumtrue
\let\Position=\centerline 

\def\Titlefont{\tenbf}   
\def\STitlefont{\tenbf}  
\def\SSTitlefont{\tensl} 
\def\Smallrm{\sevenrm}   
\def\Medrm{\tenrm}       

\newcount\EZReadMag \EZReadMag=1200
\def\EZRead{\magnification=\EZReadMag \hoffset=0pt \voffset=0pt
            \pretolerance=500 \tolerance=1000 \vbadness=1000 \hfuzz=1 true pt
            \baselineskip=1.1\baselineskip plus.1pt minus0pt
            \vbadness=2500
            \parskip=2pt plus 2pt
            \setbox\strutbox=\hbox{\vrule height .75\baselineskip
                                               depth  .25\baselineskip
                                               width 0pt}%
            \Page{5.75 true in}{8.5 true in}}
\EZRead

\font\bigit=cmti10 at 12pt
\def\scri{\hbox{\bigit i\/}}

\VersionInfo{December 19, 1996}

\Title REGULAR BLACK HOLES AND TOPOLOGY CHANGE

\Author Arvind Borde
\Email{borde@cosmos2.phy.tufts.edu}

\Address Institute of Cosmology, Department of Physics and Astronomy\\
         Tufts University, Medford, MA 02155\\
         and\\
         Department of Mathematics, Southampton College\\
         Long Island University, Southampton, NY 11968

\Abstract
The conditions are clarified under which regular (i.e., singularity-free) 
black holes can exist.  It is shown that in a large class of spacetimes 
that satisfy the weak energy condition the existence of a regular 
black hole requires topology change.
\EndAbstract


\StartPaper

Mars {\it et al.}
\StoreRef{\MMS}
\Ref{M. Mars, M.M.~Mart\'\i n-Prats and J.M.M. Senovilla, 
      \Jou{Class.\ Quant.\ Grav.}, \Vol{13}, L51 (1996).}
have recently produced
a ``Schwarzschild black hole'' that obeys the weak energy condition
\Ref{The weak energy condition requires that the energy density of matter 
be non-negative, when measured by any observer.},
but is nevertheless free of singularities.  In this letter
I clarify how the singularity-avoidance occurs, not only in their model,
but in general. My main result is that 
singularities can be avoided in a large class of black holes
only through topology change.  In proving this result I will use
methods and terminology from the ``global techniques'' area
of general relativity. Since these may be unfamiliar to many physicists,
I will give some informal definitions as they seem necessary. Precise
definitions may be found in the book by Hawking and Ellis
\StoreRef{\HE}
\Ref{S.W. Hawking and G.F.R. Ellis, \Jou{The large scale structure of
      spacetime}, Cambridge University Press, Cambridge, England (1973).}.

Regular black holes are not new.  In~1968, Bardeen
\StoreRef{\Bardeen}
\Ref{J. Bardeen, presented at GR5, Tiflis, U.S.S.R., and published
     in the conference proceedings in the U.S.S.R. (1968).}
produced a famous model, conventionally interpreted as
a counterexample to the possibility that the existence of singularities 
may be proved in black hole spacetimes without assuming
either a global Cauchy hypersurface
\Ref{A Cauchy hypersurface, $\cal S$, is a hypersurface from which the
behavior of at least a portion of the entire spacetime can be
predicted. This portion of the spacetime is called the Cauchy development 
of~$\cal S$.  If the Cauchy development is, in fact, the full spacetime,
$\cal S$ is called a global Cauchy hypersurface.}
or the strong energy condition
\Ref{The strong energy condition requires that certain combinations of the
components of the stress-energy tensor of matter be non-negative.  The condition
holds in most spacetimes of interest, but is violated, for example, in inflating 
spacetimes.}. 
(See, for example, ref.~[\HE], page~265.)  
This model is a regular black hole obeying the
weak energy condition, and it
was influential in shaping the direction of 
subsequent research
on the existence or avoidance of singularities.   
The paper in which it was originally
presented~[\Bardeen] is not readily available, but a
discussion of the model, and the role that it has played,
has since appeared in a more accessible place
\StoreRef{\Borde}
\Ref{A. Borde, \Jou{Phys. Rev. D.}, \Vol{50}, 3392 (1994).}.
In brief, the model uses the Reissner-Nordstr\"om
spacetime as inspiration.  The metric 
expressed in standard spherical coordinates
$(t,r,\theta, \phi)$ is
$$
ds^2= -f(r)\,dt^2 + {1\over f(r)}\,dr^2 + r^2\,d\theta^2
      + r^2\sin^2\theta\,d\phi^2,
$$
where Bardeen replaced the usual Reissner-Nordstr\"om function 
$$
\eqalignno{
f(r)&=1 - {2m\over r} + {e^2\over r^2}\cr
\noalign{\noindent\rm by}
f(r)&= 1 - {2mr^2\over (r^2 + e^2)^{3/2}}.\cr}
$$
When $e^2 < (16/27) m^2$ in Bardeen's model, there is an event horizon. 
The new spacetime also obeys the weak energy condition (assuming
Einstein's field equation), yet it
contains no physical singularities.
A similar example (i.e., possessing an event horizon and obeying the
weak energy condition, yet non-singular) has also been constructed
by modifying the Reissner-Nordstr\"om metric in a region near
$r=0$ and leaving it untouched outside~[\Borde\,
\Refe{A. Borde, \Jou{Singularities in Classical Spacetimes}, Ph.D.
dissertation, SUNY at Stony Brook (1982).}.

The recently proposed regular black hole~[\MMS] has a global
structure identical to these previously-discussed
Reissner-Nordstr\"om-based black holes.  
The metric here is
$$
ds^2=-{\rm e}^{4\beta}\chi\,du^2 + 2{\rm e}^{2\beta}dudr + r^2(d\theta^2 +
\sin^2\theta d\phi^2)
$$
where
$$
\eqalignno{
\chi &= 1 - {2M(r)\over r},\cr
M(r) &= \cases{m,&$r\geq 2m$\cr\cr
              {\displaystyle
              {r^3\over 8m^2}\left(10-{15r\over 2m}+{3r^2\over 2m^2}\right)},
               &$r\leq 2m$\cr},\cr
\noalign{\smallskip\noindent \rm and $\beta$ is obtained from \smallskip}
\beta'(r) &= \cases{0,&$r\geq 2m$\cr\cr
              {\displaystyle
               {5r\over 2m}\left(1-{r\over 2m}\right)^2},
               &$r\leq 2m$\cr}.\cr}
$$
This metric coincides with the Schwarzschild metric for $r\geq 2m$.
There is still an event horizon at $r=2m$ and trapped surfaces just
inside.

Yet another class of spacetimes with similar properties has been discussed by
Frolov {\it et al.}
\Refc{V.P. Frolov, M.A. Markov and V.F. Mukhanov, \Jou{Phys. Lett.~B},
\Vol{216}, 272 (1989).}
\Refe{V.P. Frolov, M.A. Markov and V.F. Mukhanov, \Jou{Phys. Rev.~D},
\Vol{41}, 383 (1990).}.
In these spacetimes, part of the region inside $r=2m$ in a  
Schwarzschild black hole is joined through a thin boundary layer
to de Sitter spacetime, allowing in some cases
singularity-avoidance to occur.
(See Barrab\`es and Frolov
\Ref{C. Barrab\`es and V.P. Frolov, \Jou{Phys. Rev.~D}
\Vol{53}, 3215 (1996).} 
and additional references cited therein.)

All these black hole spacetimes have very similar global properties.
I will therefore refer to them 
as ``Bardeen black holes.''

\StartFigure{340}{140}
   { /PtAtInf {2 0 360 arc gsave 1 setgray fill grestore stroke} def
     /RN {
          newpath
          35 0 moveto 0 35 lineto
          35 70 lineto stroke
          newpath
          105 0 moveto 140 35 lineto
          105 70 lineto stroke
          gsave
             newpath
             70 35 moveto 35 70 lineto 70 105 lineto
             105 70 lineto closepath gsave .9 setgray fill grestore
             [1 1] 0 setdash
             newpath 60 80 moveto 37 103 lineto stroke
             newpath 60 80 moveto 103 123 lineto stroke
             newpath 60 80 1.5 0 360 arc gsave 0 setgray fill grestore
          grestore
          newpath
          35 0 PtAtInf
          0 35 PtAtInf
          35 70 PtAtInf
          35 140 PtAtInf
          105 140 PtAtInf
          105 70 PtAtInf
          140 35 PtAtInf
          105 0 PtAtInf } def
       100 0 translate
       gsave
       .5 setlinewidth
       [4 4] 0 setdash
       newpath 35 70 moveto 35 140 lineto stroke
       newpath 105 70 moveto 105 140 lineto stroke
       grestore
       RN
       gsave
       1.2 setlinewidth
       newpath 35 75 moveto 55 78 85 72 105 75 curveto stroke
       newpath 3 35 moveto 40 40 85 30 137 35 curveto stroke
       grestore
   }
\label(95,105){$\matrix{\hfill r=0\cr \hbox{(origin)}\cr}$}
\label(208,105){$\matrix{r=0\hfill\cr \hbox{(origin)}\cr}$}
\label(166,79){$\cal T$}
\label(200,24){${\cal S}_1$}
\label(165,65){${\cal S}_2$}
\label(230,55){${\scri}^+$}
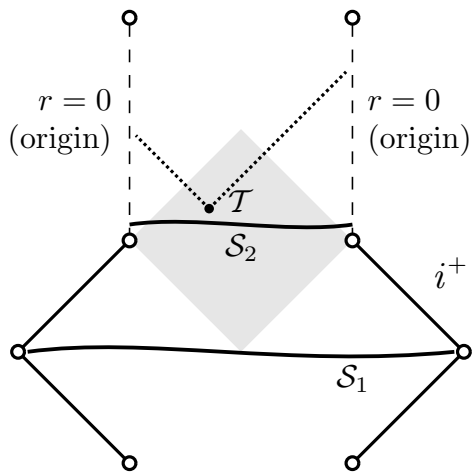
\caption{{\it The global structure of a portion of a Bardeen
black hole.}
A point in the interior of the spacetime represents a 2-sphere.
The boundaries of the diagram are drawn according to these conventions:
solid lines and hollow circles represent regions at infinity,
dashed lines represent
the origins ($r=0$) of the coordinate systems. (The two $r=0$ lines
shown above are the origins of two different coordinate patches.) 
If one imagines a series of horizontal
lines across the diagram, representing spacelike hypersurfaces, the
topology of these surfaces will switch from $S^2\times R$ in the
asymptotically flat part of the spacetime to~$S^3$ in
the region between the $r=0$ lines. For instance, the surface~${\cal S}_1$
has topology $S^2\times R$ while ${\cal S}_2$ has the topology of
a 3-sphere.
The shaded region lies inside the black hole, and contains
trapped surfaces,~$\cal T$.
One such trapped surface
is represented above by a solid dot. The dotted lines emanating
from it represent the two systems of future-directed
null rays from~$\cal T$: the ``ingoing'' and the ``outgoing.'' Each system
reaches a focal point at
$r=0$.  This light cone
has the topology of a 3-sphere.}
\EndFigure

How do Bardeen black holes avoid singularities?
Consider \Fig. It represents part of the maximally extended
spacetime of a Bardeen black hole (compare it with, for example,
fig.\thinspace3 of Mars {\it et al.}~[\MMS]).  The shaded region there
represents part of the interior of the black hole. (Signals from that
region cannot escape to future null infinity, ${\scri}^+$.)
There are trapped surfaces, $\cal T$, in this region. 
The escape from a singularity inside the black hole occurs  
because in this crucial
region where trapped surfaces exist it is possible for light rays to
``wrap around the universe;'' i.e., although both the ``ingoing'' and
``outgoing''  systems of future-directed null geodesics from $\cal T$ are
converging, the two sets converge to focii at different $r=0$
points (antipodally located with respect to each other).
The trapped surface~$\cal T$
and its future light cone $E^+({\cal T})$ both lie
in the future Cauchy development of the surface~${\cal S}_2$, and a singularity
is avoided {\sl purely because ${\cal S}_2$ is compact}. (The compactness of
${\cal S}_2$ is, in essence, what allows this spacetime to escape the
clutches of Penrose's singularity theorem of 1965
\Ref{R. Penrose, \Jou{Phys. Rev. Lett.}, \Vol{14}, 57 (1965).}; this
point is explained in detail elsewhere~[\Borde].) 

In other words, spacelike slices 
\Ref{A {\it slice\/} is an edgeless hypersurface such that no two points
on it can be connected by a timelike curve.} 
in a Bardeen black hole evolve~-- as Bardeen himself noted in the case of 
his original model~-- from a region in which they
are noncompact (e.g., ${\cal S}_1$ in \Fig), 
into a region where 
they are compact (e.g., ${\cal S}_2$): i.e., the 
Universe changes
its topology from ``open'' to ``closed.''
Such a topology-change statement must, in general, be approached with 
caution. de~Sitter spacetime, for example, contains both open and
closed spacelike slices, but can hardly be considered to 
change topology. There, however, the non-compact slices are just
poorly (or, in the case of inflationary models, deliberately)
chosen ones in that globally hyperbolic spacetime. Through every point of that
spacetime there passes a compact global Cauchy hypersurface.  
In a Bardeen black hole, that is not the case. There are regions of spacetime
through points of which no slice is compact, and other regions
through points of which every slice is compact. 

Is such topology change necessary for the existence of regular
black holes that obey the weak energy condition?   
The following theorem shows that it is:

\TheoremN
Suppose that there is a spacetime~$\cal M$ that
\item{A.} contains an eventually future-trapped surface $\cal T$,
\item{B.} obeys the null convergence condition (i.e., the Ricci tensor, $R_{ab}$,
obeys $R_{ab}N^aN^b\geq 0$ for all null vectors~$N^a$),
\item{C.} is null-geodesically complete to the future, and
\item{D.} is future causally simple, with
$E^+(X)\ne\emptyset, \forall X\subset {\cal M}$.
{\smallskip \noindent Then there is a compact slice to the causal
future of $\cal T$.}

Assumption A may be expected to hold in the interior of
a typical black hole. I am introducing here the notion of an 
{\it eventually trapped surface\/} as an extension of the usual notion
of a trapped surface. A future-trapped surface is one in which the future-directed
null geodesics that emanate orthogonally from the surface all have 
negative divergence at the surface. In an eventually future-trapped surface, the 
divergence is only required to become negative somewhere
to the future of the surface along each geodesic.
This weakening of the notion of a trapped surface is intended to cover
situations where a black hole might be growing due to infalling
matter.
Inside the black hole one may expect there to 
be closed spacelike 2-surfaces that are eventually future trapped.  
Otherwise, there would be congruences of geodesics
with non-negative divergence throughout~-- and it is hard to see how these 
could avoid escaping to infinity. It is possible to contrive situations
where geodesics inside a black hole have positive divergence, and escape to 
infinity, but to a ``different infinity'' than the one with respect to which
the black hole is defined. This happens, for example, in the $e^2=m^2$ case
of a Reissner-Nordstr\"om black hole
\Ref{I am grateful to Tom Roman for pointing this out.}. 
Such behavior does not, however, appear
to be generic, and in all reasonable cases trapped surfaces do occur
inside black holes. 
 
Assumption~B follows from the weak energy condition and a
field equation such as Einstein's. It may be replaced by a suitable
``averaged condition'' in situations where there are small violations
of the pointwise weak energy condition
\Refl{F.J. Tipler, \Jou{J. Diff. Eq.}, {\bf 30}, 165 (1978).} 
\Refn{F.J. Tipler, \Jou{Phys. Rev.~D}, {\bf 17}, 2521 (1978).} 
\Refn{G. Galloway, \Jou{Manuscripta Math.}, \Vol{35}, 209
     (1981).} 
\Refn{T. Roman, \Jou{Phys. Rev.~D}, {\bf 33}, 3526 (1986); 
     \Jou{Phys. Rev.~D}, {\bf 37}, 546 (1988).} 
\Refe{A. Borde, \Jou{Cl. and Quant. Gravity\/} \Vol{4}, 343 (1987).}.
Some form of convergence condition is crucial here.
Without it, singularities can be
avoided with no topology change
 \Refc{T. Roman and P.G. Bergmann, \Jou{Phys. Rev.~D}, {\bf 28}, 1265 (1983).}
 \Refe{D. Garfinkle, unpublished}.

Assumption~C is a weak regularity condition~-- it will hold in
any regular spacetime.

Assumption~D
may require some explanation: Let $I^+(p)$ be the
future of a point $p$ and $E^+(p)$ its future light cone.
It may be shown that $E^+(p) \subset \dot I^+(p)$.
In general, however, $E^+(p) \ne \dot I^+(p)$; i.e, the
future light cone of~$p$ is a subset of the
boundary of the future of~$p$, but is not necessarily the full boundary
of this future. A situation where this occurs has been illustrated
elsewhere~[\Borde\,%
\Refm{A. Borde and A. Vilenkin, \Jou{Phys. Rev. Lett.}, \Vol{72}, 3305 
      (1994).} 
\Refe{A. Borde and A. Vilenkin, in \Jou{Relativistic Astrophysics: 
      The Proceedings 
      of the Eighth Yukawa Symposium}, edited by M.\ Sasaki, 
      Universal Academy Press, Japan (1995).}.
Spacetimes in which
$E^+(p) = \dot I^+(p)$ for all points~$p$ are called {\it future causally
simple}. All the spacetimes in the Bardeen black hole category
discussed here are causally simple, as are the standard
Schwarzschild and Reissner-Nordstr\"om black holes and the
exterior ($r\geq 0$) Kerr black hole. 
Some form of causality assumption is necessary for the topology-change
argument of this paper to go through. If no such assumption is
made, there are scenarios possible, based on the G\"odel solution, 
that neither possess singularities nor change topology 
\Refl{My attention was drawn to this aspect of G\"odel's solution
     in 1982 by Roger Penrose.}
\Refn{A. Borde, Syracuse University report (1987); 
      Example~V in \Jou{\TeX\ by Example}, Academic Press,
      Cambridge, MA, USA (1992).}
\Refe{R.P.A.C. Newman, \Jou{Gen. Rel. and Grav.}, \Vol{21}, 981 (1989).}.

The theorem, as stated, does not directly make a statement about 
topology change. A black hole spacetime, however, usually
contains a region at infinity and may therefore be expected to
 ``start'' with a noncompact
slice~$\cal S$, with the black hole, and, hence, at least one eventually 
future-trapped surface to the future of it.  The theorem shows,
under very general conditions, that if the black hole is to be regular,
the spacetime must develop
a compact slice to the future of~$\cal S$: i.e., the topology must change
from open to closed. It follows, of course, that such a spacetime cannot 
be globally hyperbolic. There is no violation here of known results that
forbid topology change
\StoreRef{\Geroch}
\Refl{R.P. Geroch, \Jou{J. of Math. Phys.}, \Vol{8}, 782 (1967).}
\Refn{R.P. Geroch, \Jou{Singularities in the spacetime of General
 Relativity}, Ph.D. Dissertation, Princeton University (1968).}
\Refn{F.J. Tipler, \Jou{Causality Violations in General Relativity},
Ph.D. Dissertation, Univ. of Maryland (1976).}
\Refn{F.J. Tipler, \Jou{Ann. of Physics (NY)}, \Vol{108}, 1 (1977).}
\StoreRef{\Tipler}
\Refn{F.J. Tipler, \Jou{Phys. Lett.}, \Vol{165B}, 67 (1985).}
\StoreRef{\BordeTop}
\Refe{A. Borde, preprint (1996).}.
Most of these make some form of compactness assumption, ranging from
causal compactness~[\BordeTop] to more stringent 
ones~[\Geroch--\Tipler].
These assumptions do not hold in scenarios that change an open
Universe to a closed one.

\Proof
The convergence, $\theta$, of the null geodesic
generators of $E^+({\cal T})$ becomes negative, by definition, somewhere
to the future of $\cal T$ along each generator.
The quantity $\theta$ obeys~[\HE]
$$ \dot\theta \leq -{1\over 2}\theta^2 - R_{ab}N^aN^b 
              \leq -{1\over 2}\theta^2,$$ 
where assumption~B has been used in the last step.  It follows from this
and from assumption~C that
$\theta$ must diverge to~$-\infty$ at some point along each generator.
This means that each generator must 
leave $E^+({\cal T})$ within a finite affine parameter
distance~[\HE]. Thus~$E^+({\cal T})$ is compact. But, assumption~D says
that~$E^+({\cal T})=\dot I^+({\cal T})$. 
Since $\dot I^+({\cal T})$ is
the full boundary of the future of
$\cal T$, it has no edge. This is the advertised compact slice.
\EndProof

\Sectionvar Acknowledgements

I thank Tom Roman for bringing Ref.~[\MMS] to my attention, and him,
Rosanne Di\thinspace Stefano and Alex
Vilenkin for discussions.
I also thank both the Institute of Cosmology at Tufts
University and the High Energy Theory Group at Brookhaven National
Laboratory for their warm hospitality over the period when this work
was done, and the Faculty Research Awards Committee of Southampton
College for providing partial financial support.

\EndPaper